\documentclass[]{interact}
\usepackage{hyperref}
\usepackage{epigraph}
\usepackage{graphicx}
\usepackage{epstopdf}
\usepackage[caption=false]{subfig}%
\usepackage[usenames,dvipsnames]{color}
\usepackage{float}
\usepackage{tikz}
\usetikzlibrary{calc,shapes,backgrounds, arrows, positioning, fit}
\usepackage{wrapfig} 

\usepackage[natbibapa,nodoi]{apacite}
\usepackage{etoolbox}
\AtBeginEnvironment{thebibliography}{\fontsize{10}{12}\selectfont}

\theoremstyle{plain}

\theoremstyle{definition}

\theoremstyle{remark}

\begin{document}

\articletype{ORIGINAL ARTICLE}

\title{Identifying nonlinear relations among random variables: A network analytic approach}

\title{Identifying nonlinear relations among random variables: A network analytic approach}

\author{%
  Lindley R. Slipetz\textsuperscript{1} \and
  Jiaxing Qiu\textsuperscript{2} \and
  Siqi Sun\textsuperscript{1} \and
  Teague R. Henry\textsuperscript{1,2}\\[0.5em]
  \textsuperscript{1}\,Department of Psychology, University of Virginia\\
  \textsuperscript{2}\,School of Data Science, University of Virginia}

\maketitle

\begin{abstract}
Nonlinear relations, such as the curvilinear relationship between childhood trauma and resilience in patients with schizophrenia and the moderation relationship between mentalizing, and internalizing/externalizing symptoms and quality of life in youths, are more prevalent than our current methods have been able to detect. Although the use of network models has risen, network construction for the standard Gaussian graphical model depends solely upon linearity. While nonlinear models are an active field of study in psychological methodology, many models require the analyst to specify the functional form of the relation. When performing more exploratory modeling, such as with cross-sectional network psychometrics, specifying the functional form a nonlinear relation might take becomes infeasible given the number of possible relations modeled. Here, we apply a novel nonparametric approach to identifying nonlinear relations using a residualization procedure with distance correlations. We found that distance correlations excel overall at identifying nonlinear relations regardless of functional form when compared with Pearson’s and Spearman’s partial correlations and conditional mutual information. Through simulation studies and an empirical example, we show that distance correlations with residualization as a novel method can be used to identify possible nonlinear relations in psychometric networks, enabling researchers to then explore the shape of these relations with more confirmatory models.
\end{abstract}

\section{Introduction}

\epigraph{``Using a term like nonlinear science is like referring to the bulk of zoology as the study of non-elephant animals''}{- Stanislaw Ulam}

The statistical methods underlying psychological science are almost entirely built around the assumption of linear relationships between variables, from ANOVA to linear regression, to more specialized methods like structural equation modeling and even network psychometric modeling. This proliferation of linear techniques is due to efficacy, convenience and convention: linear methods capture an important aspect of any relationship between variables, the assumption of linearity makes estimation significantly simpler, and has been used effectively for well over 100+ years. While there are a suite of methods that have been developed for testing nonlinear relationships, the previous quote attributed to the mathematician Stanislaw Ulam demonstrates an issue with the use of nonlinear methods: there are an infinity of ways variables can be related nonlinearly. Yet, as psychological science moves towards the modeling of dynamical causal systems, the importance of being able to accurately detect and model nonlinear relationships, as the miss-specification of those effects in complex systems models can have much more drastic impacts on inference and interpretation than the misspecification of an effect for an associative model. In this manuscript we present a novel approach to detecting \textit{conditional} nonlinear relationships in networks of psychological variables, while controlling for the presence of confounding linear relationships. This method provides a turn-key screening method for identifying relationships that need to be further probed for nonlinear relationships, as it does not assume a specific functional form of the nonlinearity. To support the use of this method, we take a network psychometric approach to the modeling of psychological data.

Network psychometrics is a departure from the traditional ``common cause'' approach to modeling psychological phenomena \citep{Schmittmann2013}. In this latter framework, symptoms of psychological disorders are caused by some unitary underlying cause, akin to how symptoms of an infectious disease are caused by the infectious agent. Due to the theoretical issues with common cause modeling, the network psychometric approach was proposed \citep{epskamp2018networkpsychometrics, marsman2018introduction}, where a mental health condition is described as a causally interacting system of symptoms. Consider a psychological disorder like depression. In the common cause view, the symptoms of depression are ``caused'' by a latent variable that corresponds with depression severity\footnote{The issue here is, what is the causal mechanism by which this latent variable that represents depression causes symptoms? Outside of a causal explanation, the common cause model of depression is a unidimensional representation of severity, but this definition is less useful in understanding the why of depression.}, while, in the network psychometric view, depression is the whole of the causally interacting parts (e.g., having trouble sleeping causing lack of energy causing irritability \citep{Borsboom2013, Schmittmann2013, Cramer2015, Bringmann2018, Fried2020}. These differing viewpoints on disorders and how we model them has further implications on the requirements of independence within a model (i.e., local independence in the case of the common cause approach and conditional partial independence in the case of network psychometrics).

 Instead of latent variable models like confirmatory factor analysis or structural equation modeling, network psychometrics seeks to understand the relations between observed variables. The most common general analytic framework for continuous measurements is the Gaussian graphical model. In a Gaussian graphical model (which is based on partial correlations between the variables of the network, see \citet{Hallquist2019}), independence between symptoms is conditional: two variables are said to be independent after conditioning on all other variables when the partial correlation equals zero. While the Gaussian graphical model has been used productively to capture the general relationships between psychological variables, there is a core, inescapable, methodological issue to the use of partial correlation methods: these methods assume that all relationships between the observed variables are linear, which means that the the zero partial correlation is not truly statistical independence but, rather no linear relation. With the current state of the field, psychometric networks struggle to represent nonlinear relations, though some recent work has proposed a number of approaches for explicitly modeling these relations.

 The importance of being able to find nonlinear relations can be seen by the commonness of them occurring in psychological phenomena: diminishing returns for social interactions \citep{ren2021}, a curvilinear effect between age and gratitude \citep{chopik2022}, nonlinear effects of neighborhood environments on mental health \citep{zhang2022}, and the nonlinear relationship between income and mental health \citep{li2022}, among others. Furthermore, quantitative techniques have yet to catch up to these potential relations, so they are likely more prevalent than have been recorded thus far. Hence, it is vital to have an adequate method of detecting these relations.

 The purpose of this paper is to develop a general testing approach for the presence/absence of conditional \textit{nonlinear} relationships \textit{above and beyond any linear relationship} among random variables in a network psychometric setting. Our method is inspired by Generalized Covariance Modeling \citep{shah2008}, and uses a dual nonlinear/linear residualization step followed by a test of nonlinearity.   The ultimate goal is to offer a novel, broad methodology to further the applicability of network analytic methods in the psychometric context that can identify both linear and nonlinear relations using correlations that have yet to see application within psychology. The main benefit of this approach is that this nonlinearity can be detected without having to specify the functional form, a departure from previous methods. We first review current methods for psychometric network construction, including several which can capture nonlinear relationships. Following that, we present our approach, a comprehensive simulation to evaluate the sensitivity of the approach, and an empirical example of our method.

\subsection{Modeling Conditional Linear and Nonlinear Relationships}

\subsubsection{Gaussian graphical models}

A network is a data object defined as a collection of nodes representing the items to be related with one another, such as people, brain regions or behaviors, and edges representing the nature, presence and strength of those relations. In network psychometric models applied to psychopathology, a network consists of a collection of nodes representing the measured items and edges representing the statistical relationship between those items, usually as partial correlations \citep{Cramer2010, Golino2017}. 

Consider $p$ psychopathology symptom items as indicators $y_{j}$ ($j = 1,\dots, p$). Mathematically, network methods formalize the structure of a disorder as a graph, with each of the $p$ thoughts, feelings, and behaviors as nodes. The statistical relation between two variables $y_{j}$ and $y_{k}$, $k \neq j$  is signified by an edge, and the strength of this relation between each pair of variables $y_{j}$ and $y_{k}$ is symbolized by $a_{jk}$. The collection of all of these weights make up a $p \times p$ adjacency matrix, $\mathbf{Adj}$ \citep{hoffman_influence_2019}. 
	
The most commonly used framework for estimating psychometric networks is that of the Gaussian Graphical Model \citep{lauritzen_graphical_1996}. Under this framework, the adjacency matrix $\mathbf{Adj}$ is estimated as the partial correlation matrix of the variables being analyzed. Substantively, this means that the edges in the psychometric network can be interpreted as conditional linear relations between variables. While the majority of applications of the GGM make use of standard Pearson's correlations, this general modeling approach using partialled correlations can be applied using other variants such as Spearman's or poly/tetrachoric correlations. The key aspect of these approaches to note here is that the modelled relations are typically linear (monotonic in the case of Spearman's), and would therefore be unable to capture nonlinear relations that do not have a linear component.

Gaussian graphical models only directly capture linear relations, but nonlinear relations such as interactions and curvilinear relations are important to understand. For example, a curvilinear relation was found linking suicide rates and educational attainment in the elderly \citep{shah2008}, and there was evidence for a potential interaction effect between emotional dysregulation and viewing suicide as an escape \citep{al-dajani_its_2019}. The presence of nonlinear relations within psychology may be more prevalent than has been established due to a major methodological gap regarding statistical methods for detecting such relations. Current methods like moderated networks make strides into solving this, but that method specifically has problems with how interactions are specified and, thus far, there are no other available methods for detecting other curvilinear relations in networks. The proposed method is a general method that doesn't require the analyst to explicitly specify the relation or functional form.

\subsubsection{Spearman's Rank Correlation}

Spearman's rank correlation is a seemingly viable alternative for network construction with nonlinear edges and has seen some use \citep{HIROSE2017172, xue2012}. Rather than assessing the linearity of the relationship between two variables, as in Pearson's correlation, Spearman's compares the rank of each variable, assessing the monotonic relation between the two variables:

\begin{equation}
\rho_{R} = \frac{cov(R(X),R(Y))}{\sigma_{R(X)}\sigma_{R(Y)}}
\end{equation}

where $R(\cdot)$ is the rank, and $\sigma_{R(\cdot)}$ is the standard deviation of the rank. Note that $\rho_{R}$ is simply the Pearson's correlation applied to rank. A negative Spearman's correlation indicates that one variable increases while the other decreases and a positive Spearman's correlation indicates that they're monotonically similar, where a value of 1 means perfectly monotonically similar and a value of zero means there is no relationship between one's increasing and the others increasing or decreasing. Spearman's correlation has seen some application in network psychometrics \citep{isvoranu_which_2023}.

\citet{isvoranu_which_2023} assess the performance of using Spearman's correlations as input into the various GGM algorithms in order to analyze a simulated network with varying sample size (150 to 5000) and types of data. They found that, with Gaussian and ordered categorical data, they did not see an effect on the estimation, however, for skewed data, they saw improved performance in estimation. Therefore, \citet{isvoranu_which_2023} recommend Spearman's correlation for network construction as it either does not impact network estimation (in the case of Gaussian and ordered categorical data) or improves estimation (in the case of skewed data).

While Spearman's correlation captures monotonic relationships, which include nonlinear monotonic relationships, it cannot capture non-monotonic relations. Consider the simple example of the Yerkes-Dodson Law \citep{yerkes1908}, which describes a U-shaped curve between stress and performance, such that individuals perform best under moderate amounts of stress, but less well under small amounts or high amounts. Under this law, stress and performance are nonlinearly related, but they are not monotonically related. As such, if one were to apply Spearman's to this relationship, the resulting estimate of the relationship would be close to 0. 

\subsubsection{Conditional Mutual Information}

Shannon introduced mutual information in his 1948 paper, \citet{shannon_MI_1948}, as a way of quantifying how much information one random variable contains about another. Within the context of the original paper, the goal is communication; however, this method has been generalized to analyze the relation between any two random variables, including nonlinear relations. The original formula for entropy is as follows:

\begin{align}
H(Y) &= -K \sum_{i=1}^{n}p_{i}log(p_{i})
H(X, Y) &= - \sum_{i,j} p_{i,j}log(p_{i,j})
\end{align}

where K is a positive constant, $p_{i}$ is the probability of outcome $i$ occurring, $p_{i,j}$ is the joint probability of outcomes $i \& j$ occurring, and it is summed over all possible outcomes in a given variable. We use this to find the mutual information:

\begin{align}
R(X,Y) &= H(X) + H(Y) - H(X,Y)
\end{align}

Conditional mutual information was later developed, which measures the information between two variables, $X \text{ and } Y$ conditional on a third variable, $Z$. It takes the following form:

\begin{align}
R(X;Y|Z) &= H(X,Z) + H(Y,Z) - H(X,Y,Z) = H(Z)
\end{align}

Mutual information is a  candidate for identifying nonlinear relations because one do not need to specify the function form, and previous work has found success in this arena \citep[see: ][]{mandros_discovering_2020, smith_MI, steuer_mutual_2002}.

\subsubsection{Moderated Networks}

Moderation relations are examples of nonlinear relations that are of great interest to psychological science. \citet{Haslbeck2021} introduced moderated network models in which pairwise interactions can be moderated by other variables. It is closely related to the GGM, though
uses a different parameterization.
\citet{Haslbeck2021}'s model results in two estimates for pairwise interactions (i.e, $X_{n}$ regressed on $X_{m}$ and $X_{m}$ regressed on $X_{n}$) and three estimates for threeway interactions (i.e., $X_{n}$ moderates the relationship between $X_{m}$ and $X_{o}$, $X_{m}$ moderates the relationship between $X_{n}$ and $X_{o}$, and $X_{0}$ moderates the relationship between $X_{n}$ and $X_{m}$), so the authors suggest a conservative AND-rule where the mean is taken across all three values if they are all non-zero (otherwise it is set to zero). In their simulation study, they found that when the true moderator variable is known or when the true moderator is unknown but all potential moderators are included, there were consistent estimations of moderated networks in terms of high sensitivity and precision.

Even so, this faces two limitations relating to its generalizability. First, researchers would need to consider a very large space of interaction terms explicitly. For a network of size $p$, there are $\frac{p(p^{2}-p)}{2}$ possible interaction relations. So, for example, for 10 observed variables, there would be 450 potential moderated relationships. This problem arises even with the use of regularization. In addition, it only applies to nonlinear relationships which are higher-order interactions effects and requires the researcher to know the functional form of the relationship. By functional form, we mean that the researcher would at least have to know, specifically, that a moderation effect is present. However, this limitation also is beneficial in that, with moderated networks, when a researcher detects a relationship, they know it is a moderation relation.
In addition, moderated networks have the capability of fully categorical moderations, which would, in principle, allow for arbitrary functional forms to be expressed.

\subsubsection{Distance Correlations}

The common use of methods that only describe linear relations or require explicit modeling of moderating relations (as above) suggests that the field needs an approach to identify relations that are nonlinear, generally. A promising method we study here is that of distance correlations \citep{szekely2007,Szekely2014}. In distance correlations, the comparison metric relies on the Euclidean distance between the random variables being compared rather than the moments of the variables, allowing this statistic to track nonlinear relations. 

This shift from relying on covariances to using pairwise distances enables distance correlation to detect any kind of dependence no matter the functional form of the dependence. If two variables are related in any consistent way, then the distances between their observations will tend to co-vary. For example, if one variable increases  when another does (even if this relationship is not linear), the distances between similar observations in one variable will often align with distances in the other. This co-variation in the patterns of the distance relationship is what the statistic captures.

The following summary of how distance and partial distance correlations are calculated comes from \citet{Szekely2014}. Let $x_i$ and $y_i$ be 
$i \in \{1:n\}$ samples from random vector valued variables $X$ and $Y$. First, pairwise distances are calculated to construct the $n \times n$ Euclidean distance matrices ($\textbf{A}$ corresponding to $x$, and $\textbf{B}$ corresponding to $y$:

\begin{equation}
\begin{aligned}
  a_{j,k} = \parallel x_{j} - x_{k} \parallel \\
  b_{j,k} = \parallel y_{j} - y_{k} \parallel 
\end{aligned}
\end{equation}

where $\parallel \dots \parallel$ is the Euclidean norm. $\textbf{A}$ and $\textbf{B}$ summarize how different each pair of observations is from each other, across all values of $X$ and $Y$, respectively. The next step is double centering or $D$-centering, centering both the rows and columns to find the relative distance between \textbf{A} and \textbf{B}. From each element from the distance matrices, the row and column means are subtracted and the matrix mean is added, transforming $\textbf{A}$ to $\hat{\textbf{A}}$ and $\textbf{B}$ to $\hat{\textbf{B}}$. Without double centering, the variation within a variable, $X$, would influence the calculation of the covariation between $X$ and $Y$. Thus, double centering removes the influence of spread within a given variable so that all that is taken into account for the calculation of covariation between the two variables is the relationship between the two. Next, the sample distance covariance is calculated from double centered distance matrices:

\begin{align}
  distCov^{2}_{n}(X,Y) = \frac{1}{n^2}\sum_{j=1}^{n}\sum_{k=1}^{n}\hat{\textbf{A}}_{j,k}\hat{\textbf{B}}_{j,k}
\end{align}

where matrices are multiplied element-wise and the variance can be found by calculating the sample distance covariance for two identical variables. Finally, the distance correlation can be found with the following equation:

\begin{align}
  distCor^{2}(X,Y) = \frac{distCov^{2}(X,Y)}{\sqrt{distVar^{2}(X)distVar^{2}(Y)}}
\end{align}

where the correlation ranges between 0 and 1 (unlike Pearson's which can be negative). Nonzero values of the distance correlation indicate dependence of any functional form.

However, the relative magnitudes of the distance correlations are not an intuitive metric of increasing dependency. The issue here is, once we start comparing nonlinear relationships of different types, what does “increasing dependency” mean? Is $X = Y^2$  less dependent than $X = Y^3$?  We think that the use of distance correlations
 to discuss the magnitudes of relationships is problematic. Instead, we want to use this method to focus on the question of if there is a nonlinear relationship at all.

Psychometric network estimation commonly uses partial correlations, which represent the relationships between variables conditional on all other variables in the network. One may belive that the use of partial correlations in network construction allows one to understand the unique relationships between symptoms, rather than quantifying the unconditional bivariate relationships as regular correlation matrices would.

The calculation of \textit{partial} distance correlations is slightly more complicated than the calculation of distance correlations, as they rely on what is termed $U$-centering. In the case of a partial distance correlation, we are interested in the distance correlation between two vector-valued variables $X$ and $Y$, conditional on their dependence on a third vector-valued variable $Z$. We use $U$-centering in the case of partial distance correlations because $D$-centering inflates the correlation when a third variable is introduced, while $U$-centering appropriately corrects for that bias where false positives are possible on raw data. Starting from the original Euclidean distance matrices, for example $\textbf{A}$, the $U$-centered version, $\Tilde{\textbf{A}}$ is

\begin{equation}
    \Tilde{\textbf{A}} = A_{i,j} - \frac{A_{i \cdot}}{n-2} - \frac{A_{\cdot j}}{n-2} + \frac{A_{\cdot\cdot}}{(n-1)(n-2)},
\end{equation}
where $\textbf{A}_{i\cdot}, \textbf{A}_{\cdot j}, \text{ and } \textbf{A}_{\cdot\cdot}$ are the row, column and overall means respectively. Intuitively, this removes the influence of the means from the Euclidean distances, resulting only in relative distances. With the $U$-centered matrices calculated, the partial distance correlation of $X$ and $Y$, conditional on $Z$ is

\begin{equation}
    R^{*}(x,y;z) = \frac{R^{*}(x,y) - R^{*}(x,z)R^{*}(y,z)}{\sqrt{1-(R^{*}(x,z)})^{2}\sqrt{1-(R^{*}(y,z)})^{2}}
\end{equation}
where $R^*_{x,y} = \frac{(\Tilde{A}\cdot\Tilde{B})}{|\Tilde{A}||\Tilde{B}|}$, with $(\Tilde{A}\cdot\Tilde{B})$ corresponding to the normalized inner product and $|\Tilde{A}|$ corresponding to the matrix norm. This equation mimics a linear partial distance correlation equation. The partial distance correlation is a measure of conditional dependence, such that the more dependent $X$ and $Y$ are, conditional on their relation with $Z$, the higher the partial distance correlation is. However, as \citet{Szekely2014} note, for non-Gaussian variables, conditional \textit{independence} does not necessarily correspond to a partial distance correlation of 0. Rather, for non-Gaussian variables that are conditionally independent, the partial distance correlation will be close to 0, though the exact distance is not defined. The consequence of this is that significance tests based on the assumption that a partial distance correlation of 0 is a meaningful null hypothesis will be inaccurate, most likely leading to an increased rate of false positives. \citet{Szekely2014} provide a permutation based significance test for partial distance correlations that uses 0 as the null hypothesis. This is problematic as we lack set threshold for nonlinearity, and this increases tbe possibility of false positives. As distance correlations do not have the same issue (i.e., a distance correlation of 0 means independence\citep{szekely2007}), and we condition using residualization steps, we evaluate both distance correlation and partial distance correlations.

One use of distance correlations comes from the gene network reconstruction literature, where \citet{Ghanbari2018} offer a related application of distance correlations: the distance precision matrix. It is a departure from traditional Gaussian graphical models in that they have the ability to detect nonlinear relations. While mutual information and conditional mutual information have been proposed as ways of detecting nonlinearities, the authors argue that they are impractical due to requiring density estimation.

Above, we see distance correlations based on $U$-centered distance matrices (see equation 10). \citet{Ghanbari2018} use $D$-centered distance matrices (i.e., the double centered matrices used to calculate the covariance matrix), $A$, $B$, and $C$, for the random variables $X$, $Y$, and $Z$, respectively, though also assess the partial correlation determined by $U$-centering. Then, $v_{A}$, $v_{B}$, and $v_{c}$ are the vectors obtained by combining the columns of each $D$-centered matrix into a vector. Then $D$ is the matrix with all of the $D$-centered vectors as columns and the Distance Precision Matrix is the the inverse of $D^{T} \times D$.

As in the Gaussian graphical model, $p > n$ scenarios result in a singular $D^{T} \times D$ matrix, meaning that they cannot be inverted by traditional methods. \citet{Ghanbari2018} use the approach of \citet{SchäferStrimmer+2005} in which a model, $U$, of unrestricted high-dimensional parameters, $\Psi$ (where $U$ = $\hat{\Psi}$), and a model, $T$, of matching parameters of a lower-dimensional restricted submodel, $\Omega$ (where $T$ = $\hat{\Omega}$) are combined in a weighted average as 

\begin{equation}
  U^{*} = \lambda T + (1-\lambda)U
\end{equation}

where $\lambda \in [0,1]$ is the shrinkage parameter. To choose the optimal $\lambda$, the authors suggest minimizing a risk function, for example, mean squared error. This results in a regularized distance or distance precision matrix in the \citet{Ghanbari2018} simulation study. 

While the regularized Distance Precision matrix performed well (as measured by the area under the Precision-Recall curve), the success of the method is in the context of genetics, which differs from most data in psychology, particularly in terms of sample size.  Even with sample sizes of 5000, the AUPRC does not approach one for any of the methods. Given that psychology tends to have much smaller sample sizes than genetics, this is an area of exploration for future work on distance correlations on larger networks. 

\subsubsection{Conditioning via Residualization}

As we desire a test for conditional nonlinear relationships analogous to the conditional linear relationship tests of standard psychometric network construction (i.e. GGMs), one might think that methods like conditional mutual information or partial Spearman's and Pearson's correlations would be appropriate to use directly. However, these methods don't just capture the nonlinear relationship between variables, they will detect any linear relationship as well. This results in a lack of specificity, as in when a method detects both nonlinear and linear relationships, how does one identify which relationships are nonlinear, and which are purely linear?

To address this, we take a residualization approach inspired by the original Generalized Covariance Measure of conditional independence of random variables \citep[GCM;][]{Shah_2020}. There, \citet{Shah_2020} base their non-parametric dependence test on the covariances between the residualized (with respect to conditioning variables) target variables. In our approach, we propose to test the conditional nonlinear \textit{sans linear} relationship between variables $X$ and $Y$, conditional on a set of variables $\mathbf{Z}$ by first using Generalized Additive Models \citep[GAMs;][]{hastie1986generalized} to residualize $X$ and $Y$ with respect to $\mathbf{Z}$, creating $\tilde{X}$ and $\tilde{Y}$. Up to the limits of what a GAM can remove, we assume that the remaining dependency between $\tilde{X}$ and $\tilde{Y}$ is composed of nonlinear and linear components. To then remove the linear component from the relationship, we residualize $\tilde{X}$ via a simple linear regression predicting $\tilde{Y}$ from $\tilde{X}$ to create $\hat{X}$. Then we test the relationship between $\hat{X}$ and $\tilde{Y}$, which we assume contains only the nonlinear components of the relationship. We now turn to evaluating the performance of this approach across a number of data generating conditions and potential relational tests.

\section{Methods}

\subsection{Simulation Study}

The purpose of the present study is to assess the performance of our approach to identifying nonlinearly related edges, both across different conditions as well as different potential tests of the final nonlinear relationship. We performed two broad sets of simulations, evaluating a three-variable system and a four-variable system, while varying the effect strength of the target nonlinear relationship and other confounding factors. The target outcome of all simulations is the sensitivity of the tests of the conditional nonlinear relationship between variables $A$ and $C$ (the specificity of the test is also assessed).

While a full survey of potential nonlinear effect shapes is impossible, we examine several nonlinear relationships that are most applicable to psychological science: quadratic, logarithmic and multiplicative interactions. In the three-node conditions, we vary the type and strength of the target nonlinear relationship, while in the four-node condition, we fix the type of the target nonlinear relationship, and vary the type of the confounding nonlinear relationship. We evaluate the sensitivity of our approach across a number of different tests: Pearson's correlation, Spearman's correlation, conditional mutual information, distance correlations and partial distance correlations. In each case we evaluate the ability of the test to detect the nonlinear effect after the previously described residualization procedure. We also evaluate the performance of these tests on raw data, and centered data, though these results are discussed in brief and relegated to the supplementary materials. Finally, to ensure that our main condition of interest (the effect size of the nonlinear sans linear relationship between $A$ and $C$) is comparable across many different combinations of simulation conditions, we calculated the \textit{partial conditional change in $R^2$}, which is the proportion of the remaining variance that would be explained by the addition of the nonlinear effect term (when all other data generating effects are modeled). This metric is invariant to changes in the scale of the beta coefficients, the location of the estimated variables, size of the error distributions, etc.

\subsection{Data generation}

\begin{figure}[H]
    \centering
    \includegraphics[width=.7\textwidth]{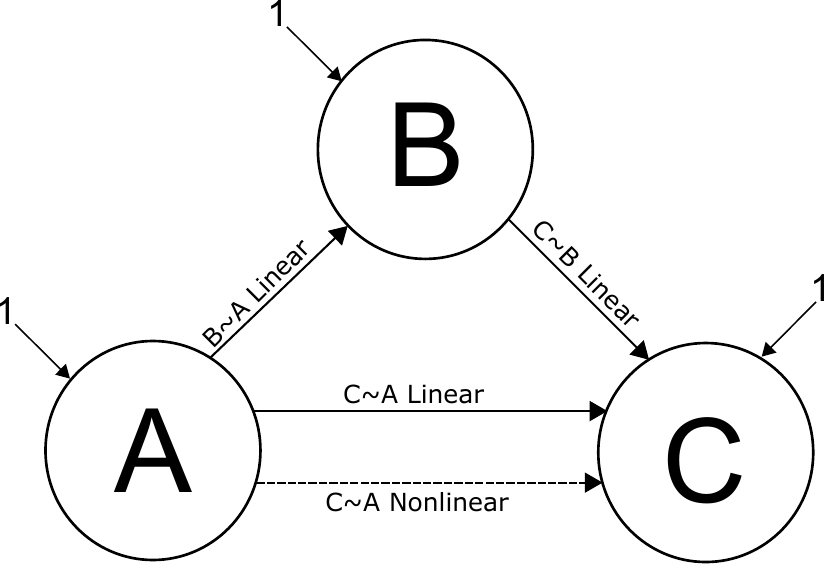}
    \caption{Schema of relations in three-node network simulations}
    \label{fig:3node}
\end{figure}

\begin{figure}[H]
    \centering
    \includegraphics[width=.7\textwidth]{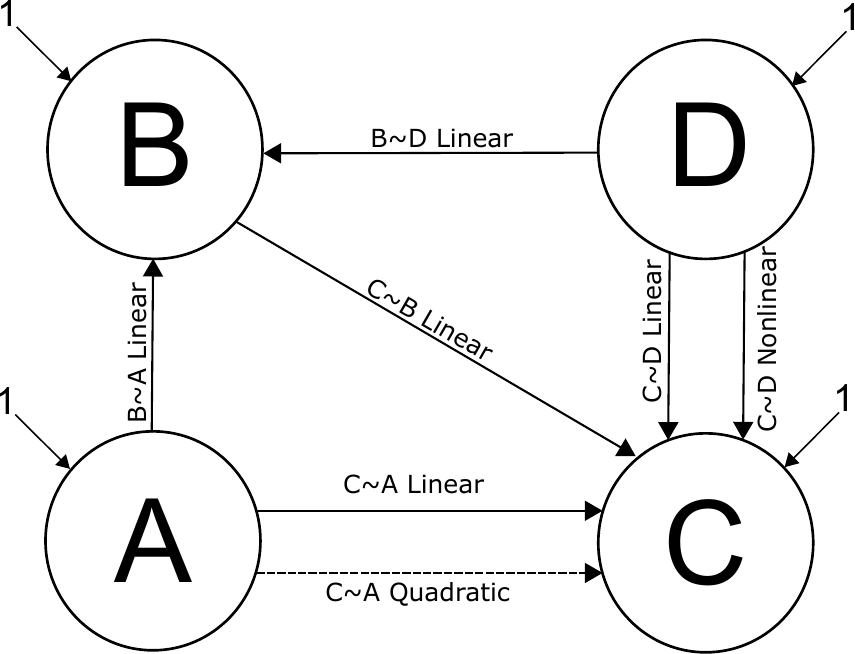}
    \caption{Schema of relations in four-node network simulations}
    \label{fig:4node}
\end{figure}

We generated data using systems of what amount to additive models with normal errors. Figures \ref{fig:3node} and \ref{fig:4node} represent the true data generating processes. In both cases, the target relationship is between $A$ and $C$.  In the three-node network (see Figure \ref{fig:3node}), Node B is a confounder that is linearly related to A and C. Node C is nonlinearly (and potentially linearly) related to A. In the case of four-node network (see Figure \ref{fig:4node}, the AC relationship is still our target, however, now the AD relationship is nonlinear and present. We specify three different types of nonlinear effects, quadratic, logarithmic, and multiplicative interaction, and systematically vary the strength of the target relationship, as well as other confounding relationships (see conditions below). Full equations for the data generating process can be found in the Supplemental Materials Section 1 and graphs of toy models can be found in Supplemental Materials Figure 1.

\subsection{Conditions}

Simulation code available is available at https://osf.io/cghpz/. As we computed the conditional partial change in $R^2$ and used it as our main factor of interest, the specific values of most parameters are less relevant than the effects presence or absence. Figures \ref{fig:3node} and \ref{fig:4node} shows graphical representations of the data generating models with the relevant parameters labeled. Directed arrows between nodes represent regression coefficients. For both the three-node and four-node simulations, sample size was set at 200 and 500. The nonlinear relationship between A and C ($C\sim A$ nonlinear and quadratic) was set at five values, resulting in a spread of partial change in $R^2$ from $\sim 0$ to $\sim .60$. In the four-node simulation, the nonlinear relationship between C and D ($C \sim D$ nonlinear) was also varied for five values, with a similar resulting spread in partial proportional change in $R^2$. All other effects were confounding factors, and were either absent or present at moderate strength. All simulation factors were fully crossed, 200 iterations were performed per cell, and within each iteration all permutation tests used 200 iterations.

We estimated distance correlation and partial distance correlations using the \texttt{energy} R package \citep{energy}, conditional mutual information with the \texttt{infotheo} R package \citep{infotheo}, and Pearson's partial correlation and Spearman's partial correlation networks both with the \texttt{ppcor} R package \citep{ppcor}. 

\subsubsection{Uncentered, centered, and residualized conditions}

In initial tests, we found that the means impacted the ability to detect nonlinear relations, so we tested three steps of handling the data. First, we tested the data as is, uncentered. Next, we assessed the methods on centered data. Finally, we residualized, removing the linearities from the relation to C, in the three- and four-node conditions, which removed the influence of the means. We implemented the residualization approach (see above).

Throughout the text, main effects correspond to purely marginal linear relationships between variables. So, in these conditions, we evaluate the performance when there is a marginal linear relationship between variables (sans moderation)  versus when there is no marginal linear relationship between variables.

\subsection{Outcomes}

The performance of the approach using different tests across different simulation factors was primarily measured by the sensitivity of the test to detect non-zero nonlinear $A \sim C$ relationships, while the specificity of the test (i.e., detecting non-zero linear relations) is also evaluated.

\section{Results: Simulation Study}

The results of the simulation demonstrate the potential of our approach to the testing of nonlinear relationships to be adapted to the network psychometric context when compared with the performance of partial Spearman's and Pearson's correlations and conditional mutual information. 

\subsection{Sensitivity for the three-node network}

\begin{figure}[H]
    \centering
    \includegraphics[width=.7\textwidth]{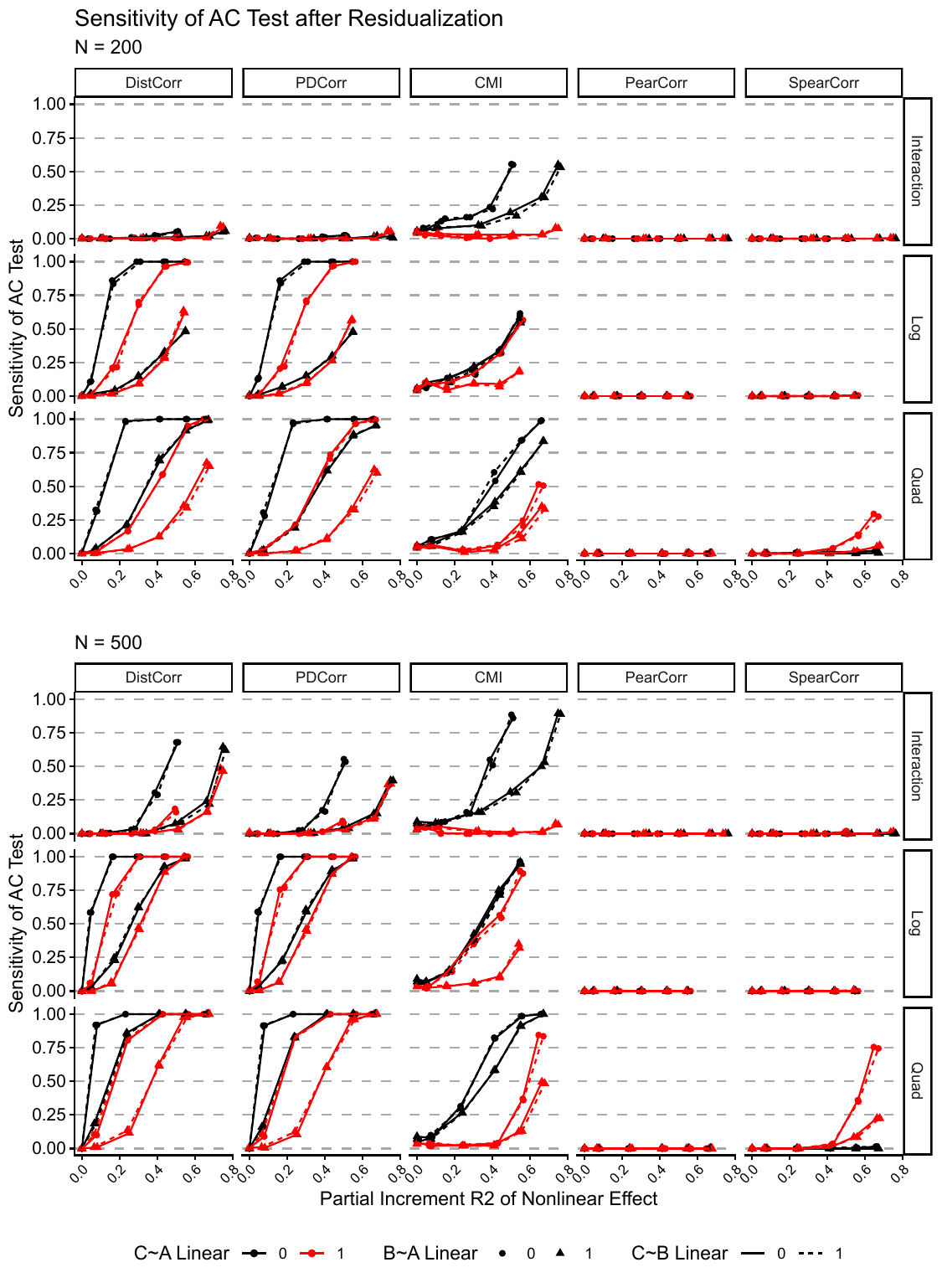}
    \caption{Sensitivity of the AC relationship in the three-node network after the residualization process. The top graph shows N = 200, while the bottom is N = 500. Increasing sample size increases sensitivity for the distance correlations and conditional mutual information.}
    \label{fig:Residual}
\end{figure}

Figure \ref{fig:Residual} shows the sensitivity for the AC edge in the three-node network after residualization (see Figures 2 and 3 in the supplementary material for graphs of the sensitivity for the AC edge for raw and centered data) when the sample size is 200 (top) and 500 (bottom). As expected, the residualization approach completely removed the linear relationship between the A and C variables, as evidenced by the 0 sensitivity shown for the Pearson's correlation across all conditions. The similar performance of Spearman's correlation, with the small exception of strong quadratic effects, is also expected, as the residualization, while not perfectly removing monotonic change, does remove the vast majority of it. 

For the three performant testing methods, distance correlations (DistCorr), partial distance correlations (PDCorr), and conditional mutual information, several common patterns were evident. First, the presence of a linear relationship between C and A (which confounds the nonlinear relationship) has a drastic negative impact on sensitivity in all cases. Second, the presence of a linear relationship between A and B had a similar negative effect on sensitivity. The presence of a linear relationship between C and B however did not seem to have any major effect on sensitivity, even conditional on other factors. 

The type of nonlinear relationship had a large effect on the sensitivity of the test. Logarithmic and quadratic effects showed similar patterns, with the negative effects of the simulation factors impacting the sensitivity of the test for quadratic relationships more than for logarithmic. The test of the multiplicative interaction term was much less sensitive than tests for the other types of nonlinear effects, controlling for the effect size, with the very best performance occurring under ideal situations (i.e. no confounding factors) and using conditional mutual information. Even then, $.80$ sensitivity is only reached when the interaction of A and B explains $\sim\%40$ of the remaining variance in $C$.

Across testing methods, the performance of the distance correlation and partial distance correlation was nearly identical, with distance correlations slightly out performing the partial distance correlation in detecting the multiplicative interaction. For the logarithmic and quadratic effects, distance correlation/partial distance correlation outperformed conditional mutual information in every condition. In the case of the multiplicative interaction however, conditional mutual information outperformed distance correlation and partial distance correlation when there was no linear relationship between C and A, but underperformed distance correlation and partial distance correlation when there was a linear relationship between C and A.

\subsection{Sensitivity for the four-node network}

\begin{figure}[H]
    \centering
    \includegraphics[width=.7\textwidth]{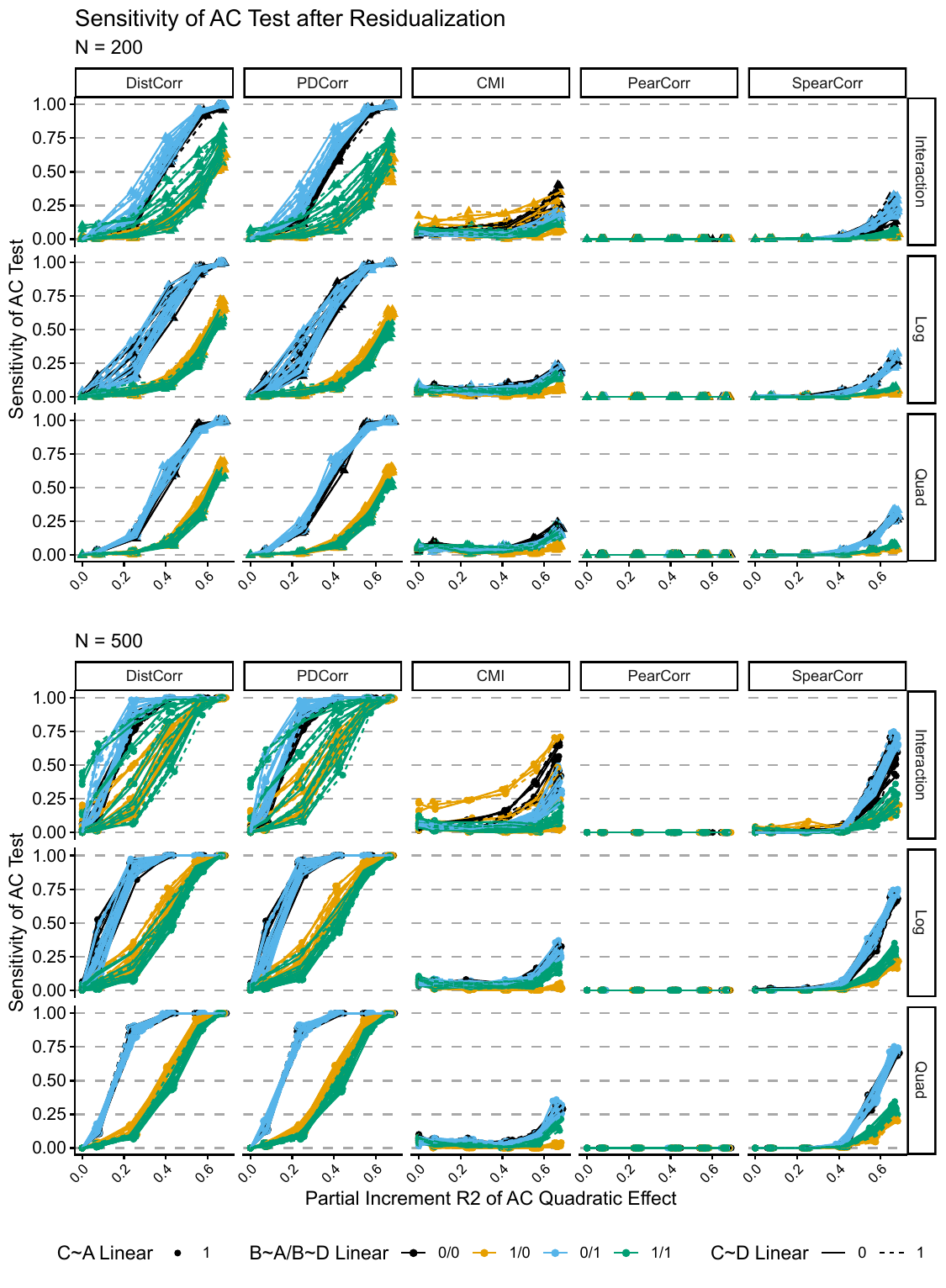}
    \caption{Sensitivity of the quadratic AC relationship in the four-node network after the residualization process. The facet on the right represents the form of the nonlinear confounder AD.}
    \label{fig:4noderes}
\end{figure}

To test the scalability of our method, we implemented a four-node network that included a confounding nonlinear AD relation, which is represented on the right facets. We examined the sensitivity for a quadratic AC relation. In Figure \ref{fig:4noderes}, notice that, for distance and partial distance correlations, the presence or absence of the linear AB relation had the most significant effect on the detection of nonlinearity. This is consistent with what we found for the three-node network. Again we see that distance and partial correlations outperform conditional mutual information and Pearson's and Spearman's partial correlations. Even with a strong nonlinear effect, while conditional mutual information with an AB relation can detect it the majority of the time, it still struggles in comparison to the proposed methods.

\subsection{Specificity}

While we did not run specific tests of specificity, we can test false positive rate when the nonlinear relationship between A and C was not present in Figures \ref{fig:Residual} . In this case, the sensitivity of the test was always close to zero, which in turn implies that specificity is high.

\section{Results: Empirical Example}

As the above demonstrates, the presented method tends to have high sensitivity, though it would be valuable to assess its capabilities on an empirical dataset. This will be done via comparison to \citet{Haslbeck2021}'s moderated networks as an evaluation of the ability to pick up the nonlinear relationships that were determined in the prior paper.

\begin{table}[]
\centering
\begin{tabular}{ccccc}
\multicolumn{1}{|c|}{Variable 1} &
  \multicolumn{1}{c|}{Variable 2} &
  \multicolumn{1}{c|}{Linear} &
  \multicolumn{1}{c|}{Moderation} &
  \multicolumn{1}{c|}{Nonlinear} \\ \hline
Hostile & Nervous   & X & X & X  \\
Hostile & Sleepy    & X & X & X \\
Hostile & Depressed & X & X & X \\
Hostile & Lonely    & X &   & X \\
Lonely  & Nervous   & X & X & X  \\
Lonely  & Sleepy    & X & X & X \\
Lonely  & Depressed & X & X & X  \\
Nervous & Sleepy    & X & X & X \\
Nervous & Depressed & X & X & X \\
Sleepy  & Depressed & X & X & X
\end{tabular}
\caption{Results from mood data for Linear (Partial Pearson's Correlation), Moderation \citep{Haslbeck2021}, and Nonlinear (Distance Correlation) relations. The linear and moderation methods were run on centered data, while the nonlinear distance method was run on residualized data The pairwise moderation relations from \citet{Haslbeck2021} would be analogous to the AB and AC relations in the interaction data generating equations in the Supplemental Materials. X signifies a significant relation.}
\label{tab:my-table}
\end{table}

\citet{Haslbeck2021} demonstrated their moderated network technique on a cross-sectional dataset of 3896 observations of the five mood variables on five point Likert scales: hostile, lonely, nervous, sleepy, and depressed. Available in the \texttt{mgm} R package, the data has been scaled to have a mean of zero and a standard deviation of one. As the data has already been centered, the comparison will between only our centered approach and these previous results. \citet{Haslbeck2021} specified that all possible moderators be included in the model.

They found nine pairwise relations (see Table 1) and four moderation relations (nervous moderating the relationship between hostile and lonely, sleepy moderating the relationship between hostile and lonely, sleepy moderating the relationship between hostile and nervous, and depressed moderating the relationship between nervous and sleepy). Pearson's correlations determined that each pairwise relation was significantly linear, so we residualized the data before performing distance correlations. Using this method, distance correlations showed greater sensitivity to the relations than the \citet{Haslbeck2021} method: the relationship between hostile and lonely was significant for distance correlations and was nonsignificant for the moderated network approach. Perhaps the distance correlation method picked up the nonlinear relationships that were not moderation relationships. The partial Pearson's correlation also picked up on these correlations with the hostile and lonely relation as well being significant, that is, all 10 pairwise relations were significant). 

Figure 3 of the Supplemental Materials depicts the relationship between the residualized hostile and lonely. Obviously, to the naked eye, a significant relationship does not immediately arise. A core issue here is that nonlinearity is nonspecific. The use of residualized distance correlations further complicates this notion as it is taking into account nonlinear conditional relationships. At the very least, what our test does is indicate to researchers that there is a relationship for which they should take a second look. Further investigation of significant nonlinear relationships is always needed to determine what kind of nonlinearity is occurring, and more importantly, is it relevant to the work at hand.
 
Therefore, \citet{Haslbeck2021} is good for exploring moderators; however partial distance correlations excel at detecting nonlinear relations. One thing to note is that distance correlations do not detect the specific interaction for an edge, rather it just identifies nonlinearity. The \citet{Haslbeck2021} method does allow for the explicit modeling of the interactions, but requires that each interaction is explicitly entered into the model.

\section{Discussion}

We ran a simulation study on three- and four-node networks to assess the ability of a novel approach to identify nonlinear edges that residualizes out the linear component of the relation and uses a distance correlation to detect the nonlinear effect. This method was compared to Pearson's and Spearman's partial correlations and conditional mutual information. We found that, for uncentered data, while distance and partial distance correlations excelled at identifying nonlinear edges and, specifically, outperformed Pearson's, Spearman's, and conditional mutual information, particularly when there is no linear AC relation present. The performance disparities between distance and partial distance correlations were further evidenced in when the residualization procedure was performed. Empirically, distance correlations and the \citet{Haslbeck2021} method picked up on different relations, perhaps indicating that the methods are more or less sensitive to difference types of nonlinear relations. This is supported by Figure \ref{fig:Residual} for distance correlations in which distance correlations are less sensitive to interactions when there is a linear AB relation for large sample sizes.

For three-node quadratic and logarithmic relations with residualized data, distance and partial distance correlations outperform  Pearson's and Spearman's partial correlations and conditional mutual information, particularly when there the AC linear relation is absent. This was further supported by the scaled four-node network. Even so, we favor the use of distance correlations over partial distance correlations as distance correlations have a meaningful zero point indicating independence while independence for partial correlations is approximately zero and unknown. This is the case even though, as Figure \ref{fig:Residual} shows, distance and partial distance correlations performed comparatively.

Alternatively, residualized conditional mutual information outperformed all other methods for interactions when there is no AC linear relation present, while struggling with quadratic and logarithmic relations. This is due to how conditional mutual information and distance and partial distance correlations incorporate the B variable within the calculation. With an interaction, the value of C is not only dependent upon the value of A but, also, B. The distance correlation method ignores B within the analysis, so it is no surprise that there would be faults in its detecting the nonlinearity. While partial distance correlations do consider the B term, it does so differently that conditional mutual information. Partial distance correlation removes the B-related aspect of the AC relation, though conditional mutual information considers the AC relation with respect to levels of B. So, it would be expected that conditional mutual information would excel over distance and partial distance correlations for this type of relation. Even so, the case where there would be no linear AC relation in the interaction would be rare, so this does not justify the use of conditional mutual information for detecting nonlinear relations. When increasing sample size, distance and partial distance correlations outperform conditional mutual information when the AC linear relation is present, suggesting our residualization process with distance correlations can be a successful tool for detecting nonlinear relations.

Our simulations show that distance correlations provide a high sensitivity and specificity test for nonlinear relationships, that this approach is agnostic to the type of nonlinearity, and that when combined with residualization, allows an analyst to disentangle the linear component of a relationship from the nonlinear components.  Overall, the decision of whether to use distance correlations for network construction lies in the hands of the researcher. This entails that distance correlations excel at identifying nonlinear relations but would be most useful in an exploratory sense, supplemented by another method like the \citet{Haslbeck2021} approach. The interrogation of nonlinear relations needs to use multiple different methods to triangulate. The distance correlation method is a high sensitivity test that leads to further more specific tests of the functional form, like \citet{Haslbeck2021}.

More work should be done to explore how the above conclusions scale to a larger, more realistic network. We demonstrated scaleability with a four-node network, but there could be limitations. Typically, when estimating larger networks, there is some way of inducing sparsity, to eliminate spurious edges. We have presented no such method here. High dimensional spaces measured with a distance metric also face the "curse of dimensionality": as the number of dimensions increases, the space becomes sparse, rendering distance metrics less useful. Finally, it becomes more computationally difficult because, for the method proposed here, you're residualizing over many variables.

Furthermore, the current study assumed cross-sectional relationships. Psychopathology, in particular, and psychological phenomena, generally, are dynamic and change over time. In addition to increasing the size of the network, future work should consider dynamic, lagged relationships between the nodes. This method could potentially be used on graphical VAR networks in which the variables of the network are correlated with the lag-1 variables. We need to test these methods for detecting nonlinear relationships over time, as psychopathology is dynamic and the relationships between variables can change over time. Thus the aim is to create a nonlinear network construction method for large scale, dynamic networks. 

\subsection{Recommendations}

Overall, we offer the following recommendations for the use of this method. It is important to residualize the data via the above procedure in order to remove the linear influences and any confounders with the following method:

\begin{enumerate}
    \item Fit Pearson's partial correlation to the data to check for significant linear influence. If significant, there is a linear element to your relation.
    \item Use a GAM to remove the influence of the confounders.
    \item Use linear residualization to remove the linear influence between the two variables you believe are nonlinearly related.
    \item Run a distance correlation on residualized data to check for nonlinear relation. If significant, there is a nonlinear relation.
    \item If the functional form of the relation is relevant, follow up with particular methods that involve functional form (e.g., \citet{Haslbeck2021} for moderation).
\end{enumerate}

Using this method in practice provides an exploratory method for researchers to detect if a nonlinear relation exists between two variables. If the goal is merely to test if nonlinearity exists, then the process ends. If the researcher wants to determine the functional form of the relation, then follow-up tests are required (e.g., \cite{Haslbeck2021}
for moderation). Overall, we provide a highly sensitive significance test procedure for nonlinear relations among random variables. 
\section{Declarations}

\begin{itemize}
    \item Funding (information that explains whether and by whom the research was supported): Not applicable
    \item Conflicts of interest/Competing interests (include appropriate disclosures): Not applicable
    \item Ethics approval (include appropriate approvals or waivers): Not applicable
    \item Consent to participate (include appropriate statements): Not applicable
    \item Consent for publication (include appropriate statements): Not applicable
    \item Availability of data and materials (data transparency): Data are open access from the mgm R package
    \item Code availability (software application or custom code): Simulation code is available at https://osf.io/cghpz/
\end{itemize}

\textit{The data come from the \texttt{mgm} R package and this simulation study was not preregistered. Simulation code is available at https://osf.io/cghpz/}

\section{Supplemental Materials: Identifying nonlinear relations among random variables: A network analytic approach}

\subsection{Data generating process}

The following are the equations for three-node networks. For quadratic relations, the variables are defined as follows:

\begin{align}
A &\sim N(\mu_A, \sigma_A)\\
B &= \beta_{ab}A + \epsilon\\
\epsilon &\sim N(\mu_B, \sigma_B)\\
C &= \beta_{non}A^{2} + \beta_{lin}A + \beta_{con}B + \nu\\
\nu &\sim N(\mu_C, \sigma_C)
\end{align}

For interaction relations, the variables are defined as follows:

\begin{align}
A &\sim N(\mu_A, \sigma_A)\\
B &= \beta_{ab}A + \epsilon\\
\epsilon &\sim N(\mu_B, \sigma_B)\\
C &= \beta_{non}AB + \beta_{lin}A + \beta_{con}B + \nu\\
\nu &\sim N(\mu_C, \sigma_C)
\end{align}

For logarithmic relations, the variables are defined as follows:

\begin{align}
A &\sim N(\mu_A, \sigma_A)\\
B &= \beta_{ab}A + \epsilon\\
\epsilon &\sim N(\mu_B, \sigma_B)\\
C &= \beta_{non}log|A| + \beta_{lin}A + \beta_{con}B + \nu\\
\nu &\sim N(\mu_C, \sigma_C)
\end{align}

In the four-node network, D is added as a nonlinear confounder. The variables are defined as follows in the four node network with a quadratic AC relationship:

If AD is quadratic:
\begin{align}
A &\sim N(\mu_A, \sigma_A)\\
D &= \beta_{ad}*A^2 + \gamma\\
B &= \beta_{ab}*A + \epsilon\\
C &= \beta_{non}*A^2 + \beta_{lin}*A + \beta_{con}*B + \beta_{con2}*D + \nu\\
\gamma &\sim N(\mu_D, \sigma_D)\\
\epsilon &\sim N(\mu_B, \sigma_B)\\
\nu &\sim N(\mu_C, \sigma_C)\\
\end{align}

If AD is an interaction:
\begin{align}
A &\sim N(\mu_A, \sigma_A)\\
B &= \beta_{ab}*A + \epsilon\\
D &= \beta_{ad}*A*B + \gamma\\
C &= \beta_{non}*A^2 + \beta_{non2}*AD + \beta_{lin}*A + \beta_{con}*B + \beta_{con2}*D + \nu\\
\gamma &\sim N(\mu_D, \sigma_D)\\
\epsilon &\sim N(\mu_B, \sigma_B)\\
\nu &\sim N(\mu_C, \sigma_C)\\
\end{align}

If AD is a logarithmic:
\begin{align}
A &\sim N(\mu_A, \sigma_A)\\
B &= \beta_{ab}*A + \epsilon\\
D &= \beta_{ad}*log|A| + \gamma\\
C &= \beta_{non}*A^2 + \beta_{lin}*A + \beta_{con}*B + \beta_{con2}*D + \nu\\
\gamma &\sim N(\mu_D, \sigma_D)\\
\epsilon &\sim N(\mu_B, \sigma_B)\\
\nu &\sim N(\mu_C, \sigma_C)\\
\end{align}

\subsection{Toy models of nonlinear relations}

\begin{figure}[H]
    \centering
    \includegraphics[width=.8\textwidth]{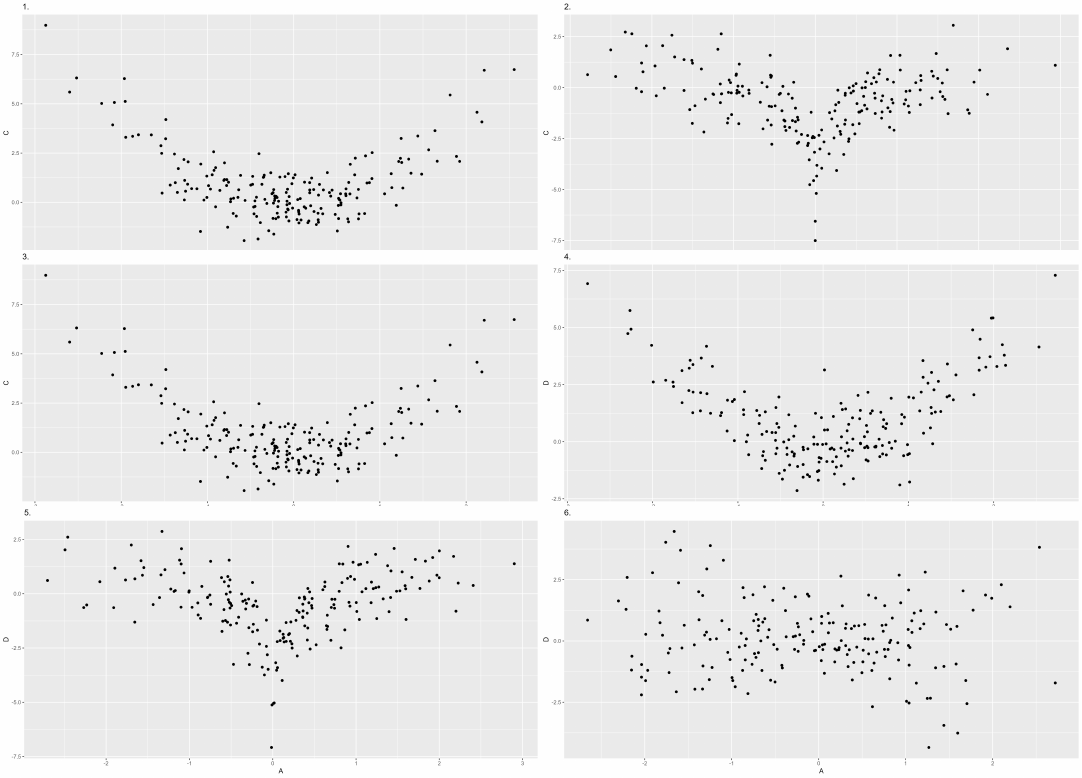}
    \caption{Example graphs of nonlinear relations. 1: 3 node quadratic; 2; 3 node logarithmic; 3: 3 node interactions; 4: 4 node quadratic; 5: 4 node logarithmic; 6: 4 node interaction.}
    
\end{figure}

\subsection{Graph of sensitivity for raw and centered data for the three-node network}

\begin{figure}[H]
    \centering
    \includegraphics[width=.8\textwidth]{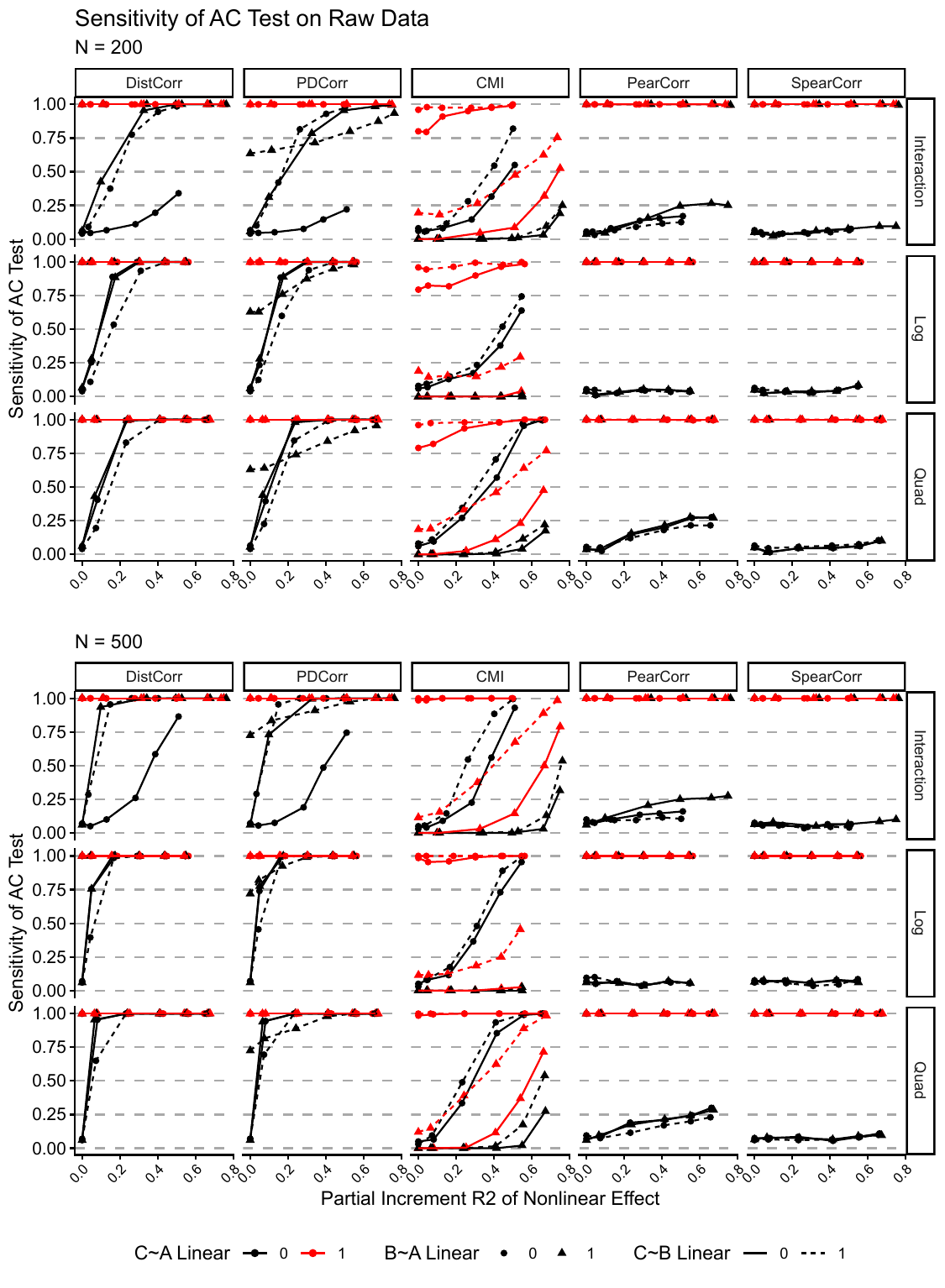}
    \caption{Sensitivity for raw data AC edge for sample sizes of 200 and 500 for the three-node network.}
    \label{fig:raw}
\end{figure}

\begin{figure}[H]
    \centering
    \includegraphics[width=.8\textwidth]{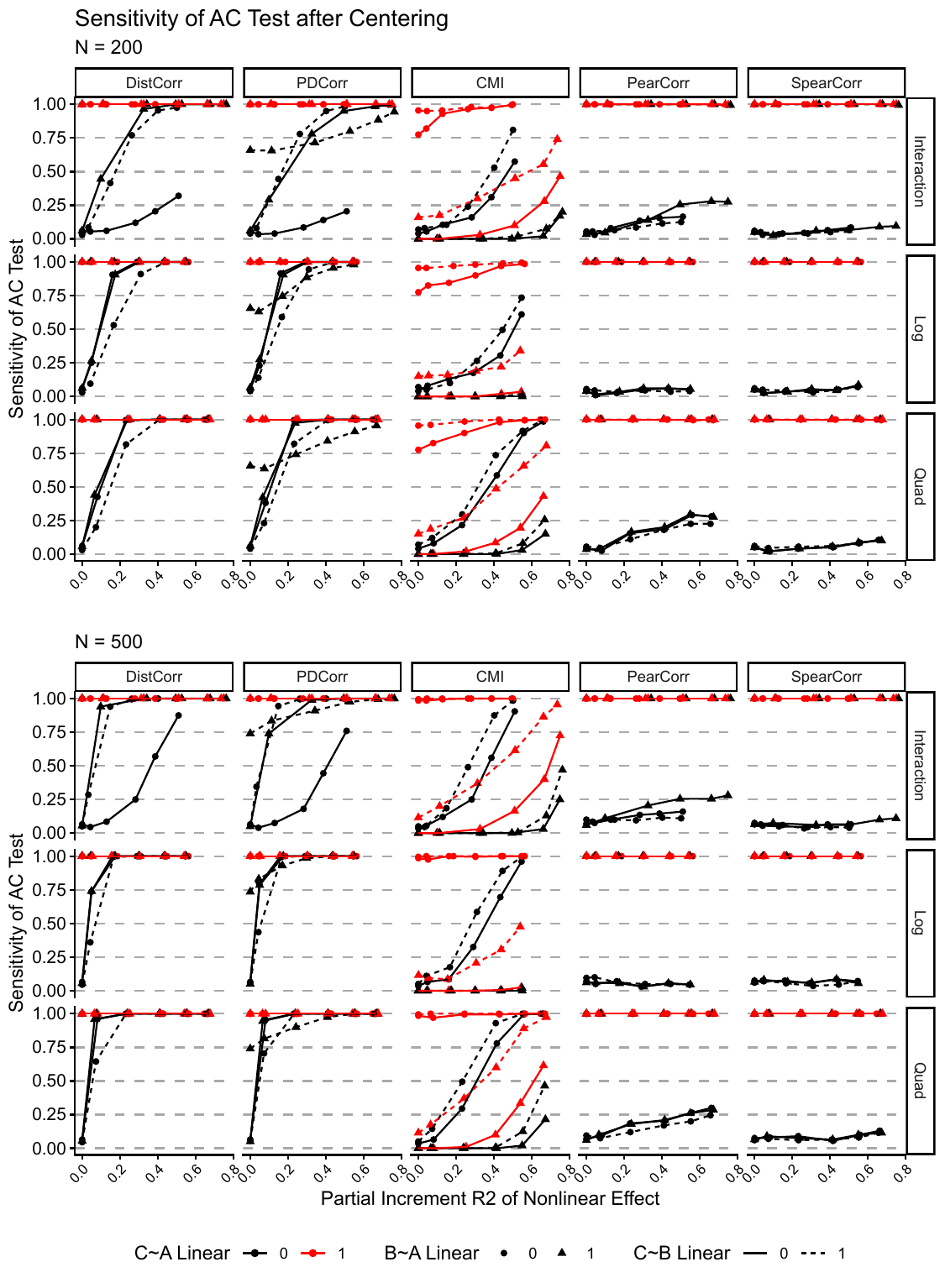}
    \caption{Sensitivity for centered data AC edge for sample sizes of 200 and 500 for the three-node network}
    
\end{figure}

\subsection{Graphs of sensitivity for raw and centered data for the four-node network}

\begin{figure}[H]
    \centering
    \includegraphics[width=.8\textwidth]{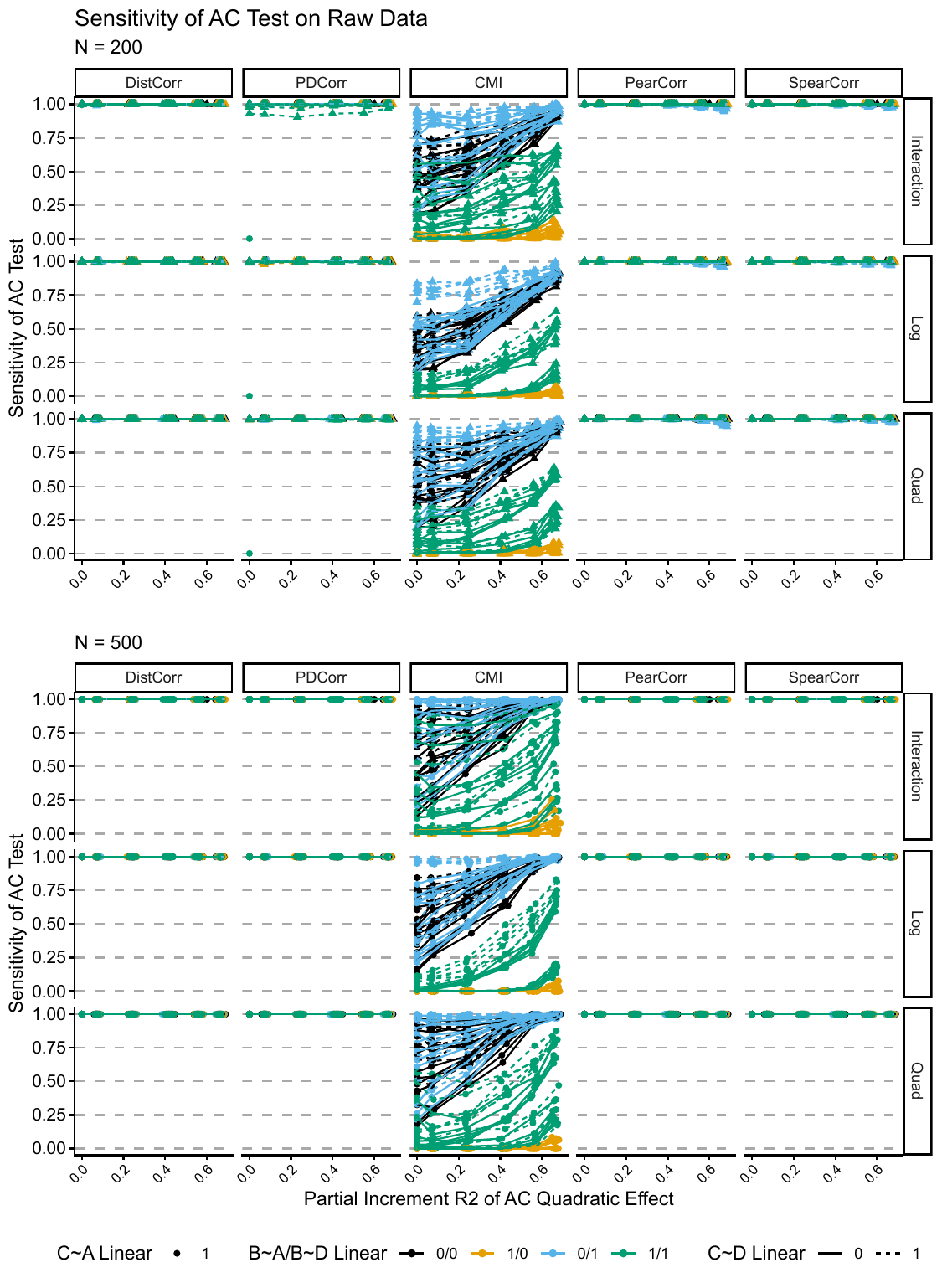}
    \caption{Sensitivity for the raw data quadratic AC edge for sample sizes of 200 and 500 for the four-node network. The facet on the right represents the form of the nonlinear confounder AD.}
    
\end{figure}

\begin{figure}[H]
    \centering
    \includegraphics[width=.8\textwidth]{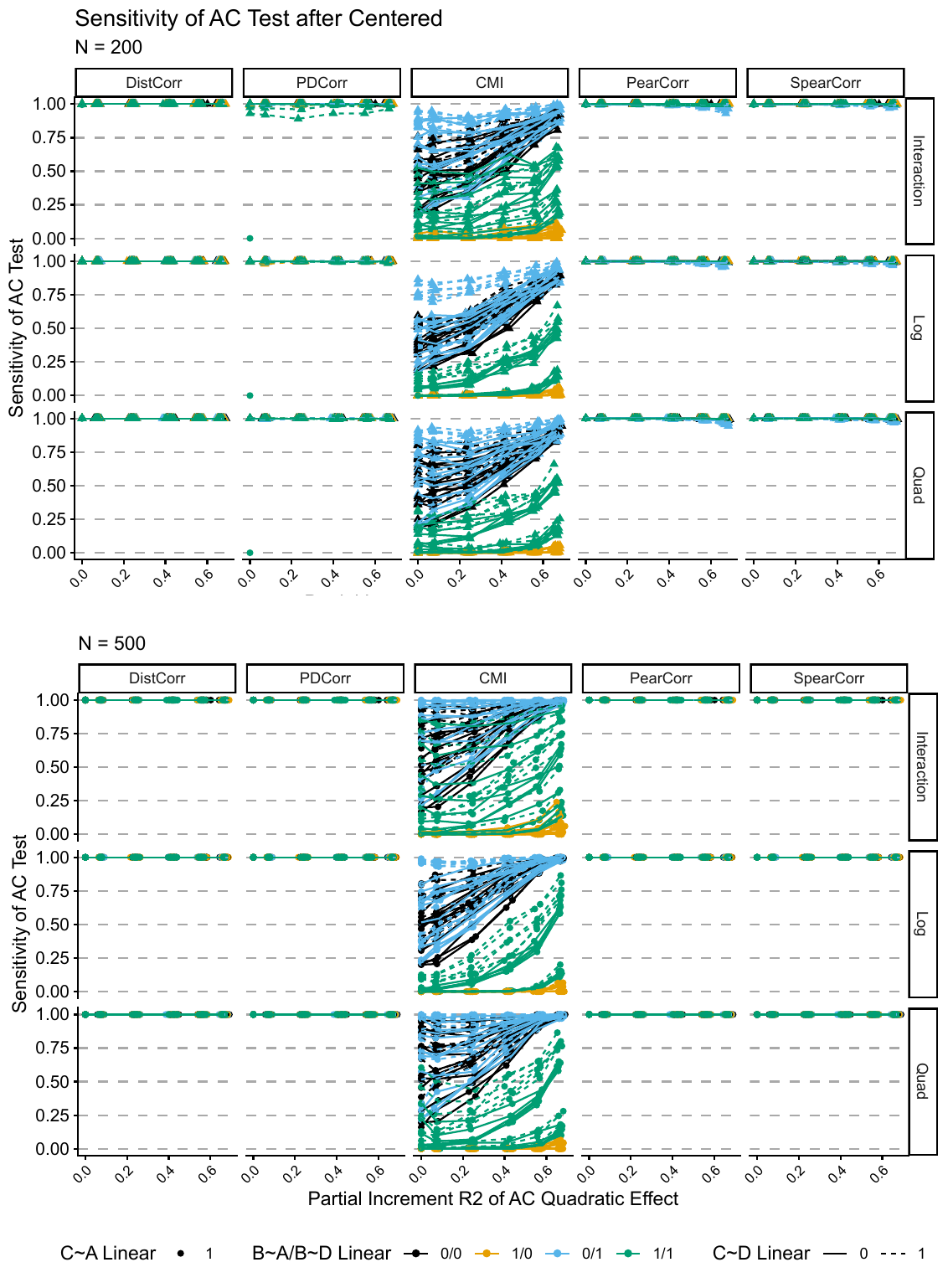}
    \caption{Sensitivity for the centered data quadratic AC edge for sample sizes of 200 and 500 for the four-node network. The facet on the right represents the form of the nonlinear confounder AD.}
    
\end{figure}

\subsection{Lonely versus Hostile empirical example}

\begin{figure}[H]
    \centering
    \includegraphics[width=.8\textwidth]{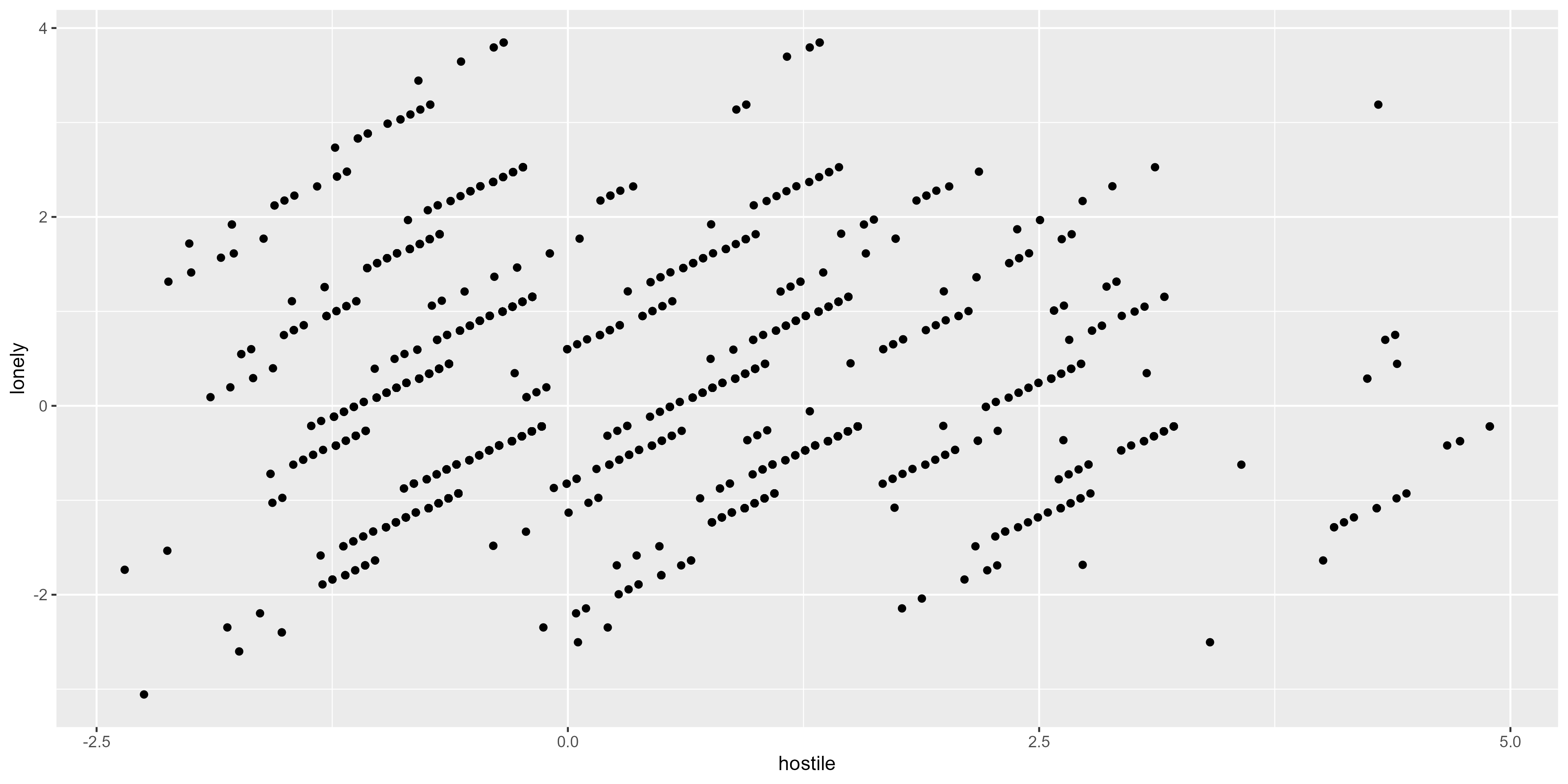}
    \caption{The relationship between lonely and hostile after residualizing.}
    
\end{figure}

\bibliography{Paper1Bib.bib}
\bibliographystyle{apacite}

\end{document}